\documentclass[conference]{IEEEtran}
\IEEEoverridecommandlockouts
\usepackage{cite}
\usepackage{balance}
\usepackage{amsmath,amssymb,amsfonts}
\usepackage{algorithmic}
\usepackage{graphicx}
\usepackage{textcomp}
\usepackage{comment}
\usepackage{xcolor}
 \usepackage{booktabs}

\usepackage{caption}
\usepackage{cleveref}
\usepackage{subcaption}
\usepackage{tikz}

\def\BibTeX{{\rm B\kern-.05em{\sc i\kern-.025em b}\kern-.08em
    T\kern-.1667em\lower.7ex\hbox{E}\kern-.125emX}}
\begin{document}

\title{
ML-based Approaches for Wireless NLOS Localization: Input Representations and Uncertainty Estimation 
}

\author{
\IEEEauthorblockN{
Rafayel Darbinyan\IEEEauthorrefmark{1}\IEEEauthorrefmark{2}, 
Hrant Khachatrian\IEEEauthorrefmark{1}\IEEEauthorrefmark{2}, 
Rafayel Mkrtchyan\IEEEauthorrefmark{1}\IEEEauthorrefmark{2}, 
Theofanis P. Raptis\IEEEauthorrefmark{3}
}
\IEEEauthorblockA{
\IEEEauthorblockA{\IEEEauthorrefmark{1}Yerevan State University, Yerevan, Armenia. Email: \{r.darbinyan, hrant.khachatrian, rafayel.mkrtchyan\}@ysu.am}
\IEEEauthorblockA{\IEEEauthorrefmark{2}YerevaNN, Yerevan, Armenia}
\IEEEauthorrefmark{3}Institute of Informatics and Telematics, National Research Council, Pisa, Italy. Email: theofanis.raptis@iit.cnr.it
}
}

\maketitle
\begin{tikzpicture}[remember picture,overlay]
\node[anchor=south,yshift=10pt] at (current page.south) {\fbox{\parbox{\dimexpr\textwidth-\fboxsep-\fboxrule\relax}{
  \footnotesize{
    This work has been submitted to the IEEE for possible publication. Copyright may be transferred without notice, after which this version may no longer be accessible.
  }
}}};
\end{tikzpicture}

\begin{abstract}
The challenging problem of non-line-of-sight (NLOS) localization is critical for many wireless networking applications. The lack of available datasets has made NLOS localization difficult to tackle with ML-driven methods, but recent developments in synthetic dataset generation have provided new opportunities for research. This paper explores three different input representations: (i) single wireless radio path features, (ii) wireless radio link features (multi-path), and (iii) image-based representations. Inspired by the two latter new representations, we design two convolutional neural networks (CNNs) and we demonstrate that, although not significantly improving the NLOS localization performance, they are able to support richer prediction outputs, thus allowing deeper analysis of the predictions. In particular, the richer outputs enable reliable identification of non-trustworthy predictions and support the prediction of the top-$K$ candidate locations for a given instance. We also measure how the availability of various features (such as angles of signal departure and arrival) affects the model's performance, providing insights about the types of data that should be collected for enhanced NLOS localization. Our insights motivate future work on building more efficient neural architectures and input representations for improved NLOS localization performance, along with additional useful application features.
\end{abstract}


\section{Introduction}

A traditional problem in networked environments is the wireless source node localization, in which the location of wireless transmitter has to be estimated from a given set of noisy measurements. Finding the source node's location may be necessary in various use case applications, such as industrial networking \cite{9039732}, emergency response \cite{Papaioannou2021}, smart agriculture \cite{ANGELOPOULOS2020107039} \cite{Chiu_2020_CVPR}, mobile crowdsensing systems \cite{ANGELOPOULOS201595}, and asset tracking \cite{https://doi.org/10.48550/arxiv.2006.05397}. A localization algorithm's objective is to analyze network and environmental parameters, in order to precisely determine the location of the source node. Conventional methods often employ a number of nodes at various predetermined locations in order to perform localization. Utilizing triangulation or trilateration, the geographical data are utilized to estimate the source's location \cite{7928257}. When there is no possibility of employing multiple nodes, however, the localization task becomes much more challenging. For example, when a single base station needs to localize one source node with a line-of-sight (LOS) radio link, usually wireless signal parameters are used \cite{7849233}.

A large part of the state-of-the-art adopts a joint ``time/direction of arrival'' localization strategy. However, when multipath (radio signals reaching the receiving antenna by two or more radio paths) is present, such strategies can drastically fail \cite{4384492}. To suppress the non-line-of-sight (NLOS) radio links, several strategies have been devised. They frequently consider NLOS pathways as noise, despite the fact that many of them use statistical models of the transmission channel \cite{6882333}. 

In the recent literature, various machine learning (ML) methods have been successfully employed to address selected wireless radio link problems \cite{8743390}. Results on prediction of spectrum \cite{8428623}, path loss \cite{9354041}, throughput \cite{9860971}, channel state information \cite{8653275} and received power \cite{9026781} demonstrate properly designed and trained ML methods can lead to very accurate and computationally efficient improvements.

The ML-assisted progress on localization problems though is less vibrant. This happens because of two main research bottlenecks \cite{8726079}: (i) ML-based localization needs large-scale high-accuracy location data as support. Currently, the collection, transmission, and storage of such massive amounts of location data face great challenges, and, (ii) the observed wireless signal characteristics are significantly influenced by environmental alterations, diverse network densities, weather conditions and obstacles. Therefore, the majority of the current wireless ML research relies on datasets produced using statistical models in constrained contexts. The statistical data prevent the trained ML models from being further tuned for a particular situation, and the constrained surroundings reduce the generalization potential \cite{Galstyan_2022_CVPR} of the learned ML models. 

Very recently, the reliable generation and use of synthetic wireless data was suggested as a way to partially tackle this issue \cite{WAIR-D,https://doi.org/10.48550/arxiv.2006.05397}. It is clear that such datasets can assist the design and evaluation ML methods for localization tasks. The most important features of such datasets are (i) the detailed radio path parameters (typically through the ray-tracing simulator) with given environment settings, (ii) the ability to support both dense and sparse deployments of varying carrier frequencies, promoting diversity and generalization, and, (iii) the map generation from the real world settings with urban or rural layout information.

\textbf{Our contribution:} In this paper, we investigate the problem of single-source, single-base station localization, with a particular focus on exploring interesting aspects of the challenging NLOS case. Specifically, we explore three different input representations, including single wireless radio path features as vector inputs, wireless radio link multi-path features as matrix inputs, and image-based input representations. While we borrow and use as benchmark the first approach from a dataset's codebase, we design and implement two new approaches for the two latter cases of input representation. Through a comparative analysis we uncover the performance trade-offs of each approach and we demonstrate that, surprisingly, these more fine-grained input representations do not improve the NLOS localization performance efficiency. However, given that the fine-grained approaches support richer prediction outputs enabling deeper visual analysis of the predictions, we proceed to identify important features in the input data and to show how their availability affects the prediction performance. Finally, we provide various insights into the scattered outputs, which can be used to identify incorrect predictions. Overall, our methodology is evaluated via a recently published large-scale synthetic wireless signal dataset, and, to the best of our knowledge, for the first time in the state-of-the-art contributes to expanding our understanding of the impact of diverse input representations on the complex problem of NLOS localization.

The roadmap of the paper is as follows: In section \ref{sec:related}, we briefly conduct a literature review and we position our contribution within the realm of the already published works. In section \ref{sec:problem}, we provide the problem definition and we model all the necessary parameters for the methodological design of the following section. In section \ref{sec:models}, we present three fundamentally different in terms of input representation ML-driven wireless localization approaches, two out of which are designed by us and one borrowed from the codebase of the dataset \cite{WAIR-D}. In section \ref{sec:perf}, we evaluate the performance of the presented approaches on the synthetic dataset \cite{WAIR-D}. We focus on highlighting diverse insights acquired by exploring the different input representation potentials, including the abilities to predict top K location candidates and to detect uncertain predictions. In section \ref{sec:conc}, we conclude the paper.



\section{Related works} \label{sec:related}

Wireless localization (sometimes also referred to as wireless positioning) has been a long studied topic in the wireless networks and communications literature when addressed with analytical methods. ML-driven wireless localization is a relatively new area of research, mainly because of the lack of suitable large-scale training datasets. The lack of datasets is due to the particular nature of the application domain of wireless networking, in which data collection involves fine-grained hardware configurations, where a priori knowledge, such as a special channel impulse response, probability density functions of the received signal parameters and so on, is hard to acquire in practice.

In the scope of our paper, we focus on the application field of outdoor localization. Therefore, indoor localization requirements and constraints, such as the ones presented in \cite{doi:10.1080/17489725.2020.1817582}, \cite{9264122} and \cite{7577201} lie beyond the scope of the paper. In the field of outdoor localization, there has been a large body of works based on (non-ML-driven) analytical methods, taking into account not only temporal (for example, time-of-arrival: ToA), but also geometric characteristics (for example angle-of-departure and angle-of-arrival: AoD and AoA) of the wireless signal transmission and reception. Particular emphasis has been given to tackling the NLOS case of wireless localization, which is also the core motivation for our paper. Specifically, in the NLOS case, the majority of the analytical methods in the literature, such as \cite{9443804}, \cite{4850277}, \cite{Wu2022} and \cite{6854860}, assume knowledge on ToA, AoD and AoA. Some works assume more limited knowledge only on, for example ToA and AoD \cite{6843297}, or ToA and AoA \cite{4384492}, \cite{9843909}. Some other works use some assistance from employing also LOS elements \cite{8904260}, \cite{8304042}.

The body of works on ML-driven methods present some interesting first attempts on addressing various wireless localization challenges. Leaving aside the works on indoor localization, such as \cite{10012991}, \cite{8555654}, \cite{9769439}, \cite{9681832} and \cite{9620795}, some works have investigated how to use ML in order to solve parallel problems related to outdoor localization, such as AoA estimation \cite{9682146}, NLOS ranging error \cite{10000807} and LOS/NLOS identification \cite{9411236}, \cite{8947166}, \cite{6820422}. On the specific problem of wireless localization, the existing works focus mainly on different network models and capabilities, such as mobile networks in \cite{9834010}. Therefore, to the best of our knowledge, this is the first work on investigating crucial aspects of the wireless outdoor localization process via ML-driven approaches. Unlike the mentioned works, in this paper, we utilize a large-scale (synthetic) wireless propagations dataset.

\section{Problem formal definition} \label{sec:problem}


In an outdoor area $u_l$ (also referred to as map), with $l=1, \ldots, L$, where $L$ is the total number of such available areas, consisting of open space and obstacles (such as buildings), we have a set $V_l$ of user equipments (also referred to as source nodes or transmitters) and a set $W_l$ of base stations (also referred to as sinks or receivers). Given a source node $i \in V_l$ situated in an undisclosed location $v_i \in \mathbb{R}^2$ and a base station $j \in W_l$ situated in a known location $w_j \in \mathbb{R}^2$, the problem is to estimate the location of the source node $v_i$ based on available radio propagation parameters. For each pair of $i \in V_l$ and $j \in W_l$ we define a single radio link as a combination of $K_{i,j}$ distinct radio paths with the following properties:

\begin{itemize}
    \item Path delays: the time it takes for the signal to travel from the source node to the base station through all paths, denoted by $\tau_{i,j}$ $\in \mathbb{R}^{K_{i,j}}$.
    \item Angles of Arrival (AoA): the direction of the signals as they arrive at the base station, denoted by $\psi_{i,j}$ $\in \mathbb{R}^{K_{i,j}}$.

    \item Angles of Departure (AoD): the direction of the signals as they depart from the source node, denoted by $\phi_{i,j}$ $\in \mathbb{R}^{K_{i,j}}$.


\end{itemize}

Note that $\tau_{i,j}$ ($\phi_{i,j}$, $\psi_{i,j}$) is a $K_{i,j}$ dimensional vector as it holds one number per path. We use the notation $\tau_{i,j,k}$ to specify the value corresponding to the $k$-th path.

The goal is to find such estimator $f$ that minimizes
$$
\sum_{l=1}^{L}\sum_{i=1}^{V_l} \sum_{j=1}^{W_l} \lVert {f(u_l, w_j, \tau_{i,j}, \psi_{i,j}, \phi_{i,j})-v_i} \rVert 
$$

where the first sum is over urban areas (maps), the second and third sums are over source nodes and base stations, respectively.


\begin{figure}[t!]
    \centering
    \includegraphics[width=\columnwidth]{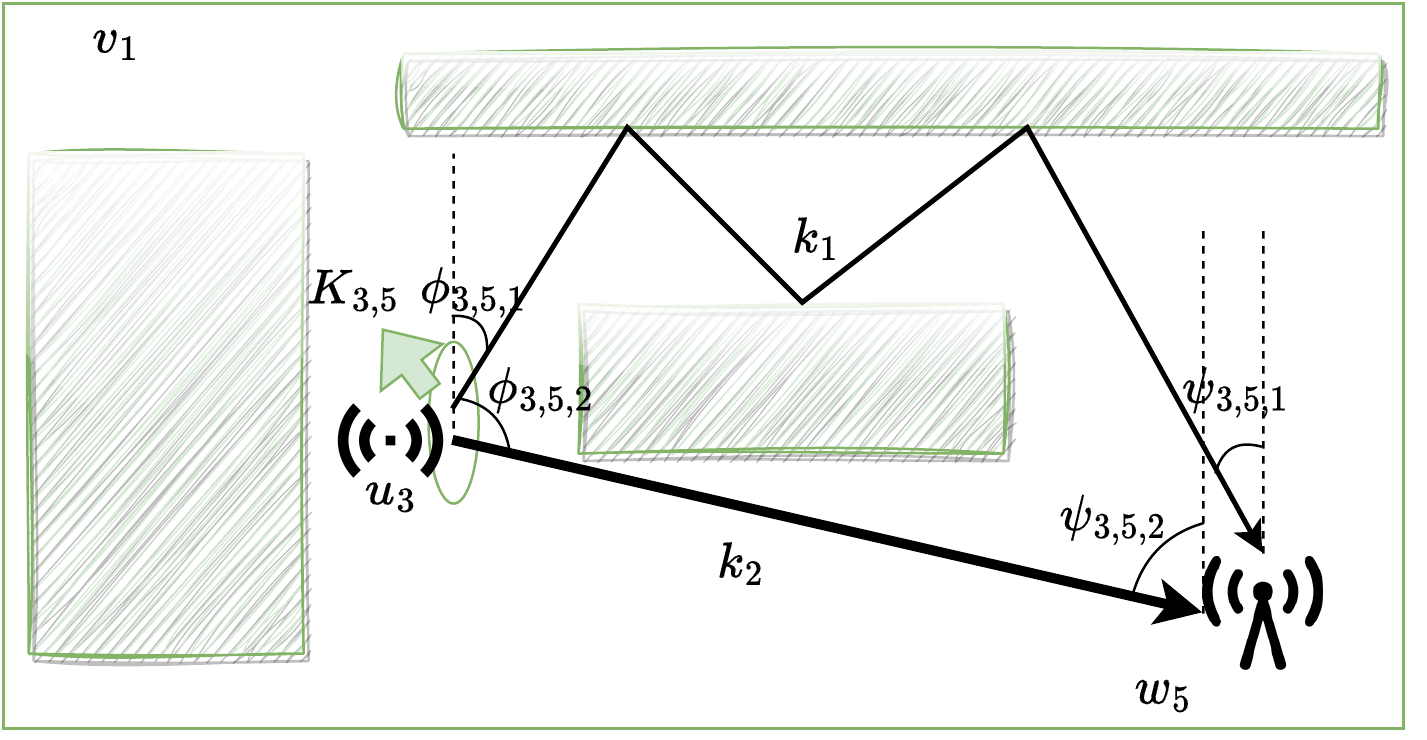}
    \caption{An example instance of the problem, with $l = 1, i = 3, j = 5, K_{3,5} = 2$.}
    \label{fig:example}
\end{figure}

\section{ML-driven approaches} \label{sec:models}

\begin{figure*}[t!]
     \centering
     \begin{subfigure}[b]{0.19\textwidth}
         \centering
         \includegraphics[width=\textwidth,trim={3cm 1cm 3cm 1cm},clip]{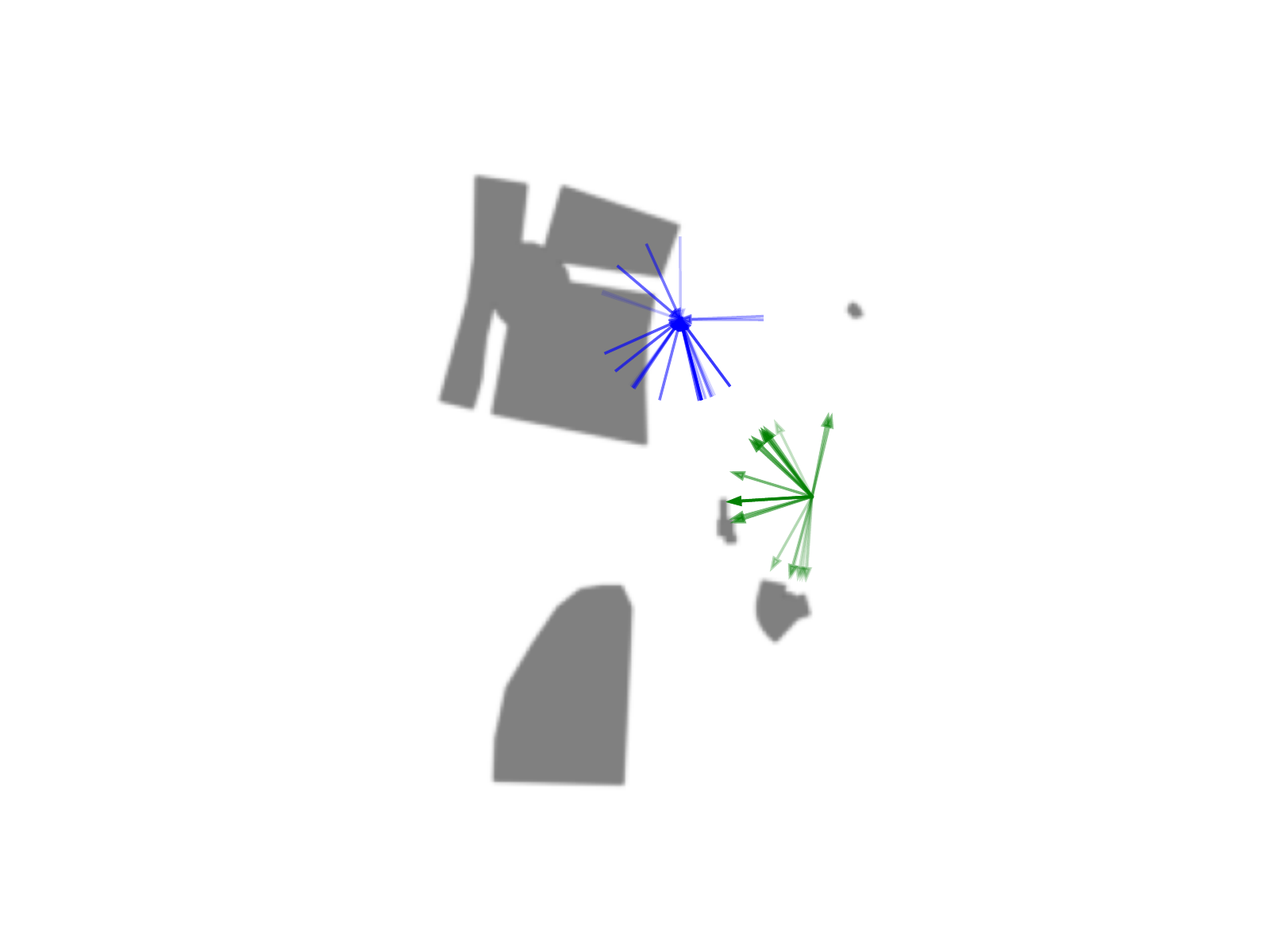}
         \caption{Original data}
         \label{fig:input-sample}
     \end{subfigure}
     \hfill
     \begin{subfigure}[b]{0.19\textwidth}
         \centering
         \includegraphics[ width=\textwidth,trim={3cm 1cm 3cm 1cm},clip]{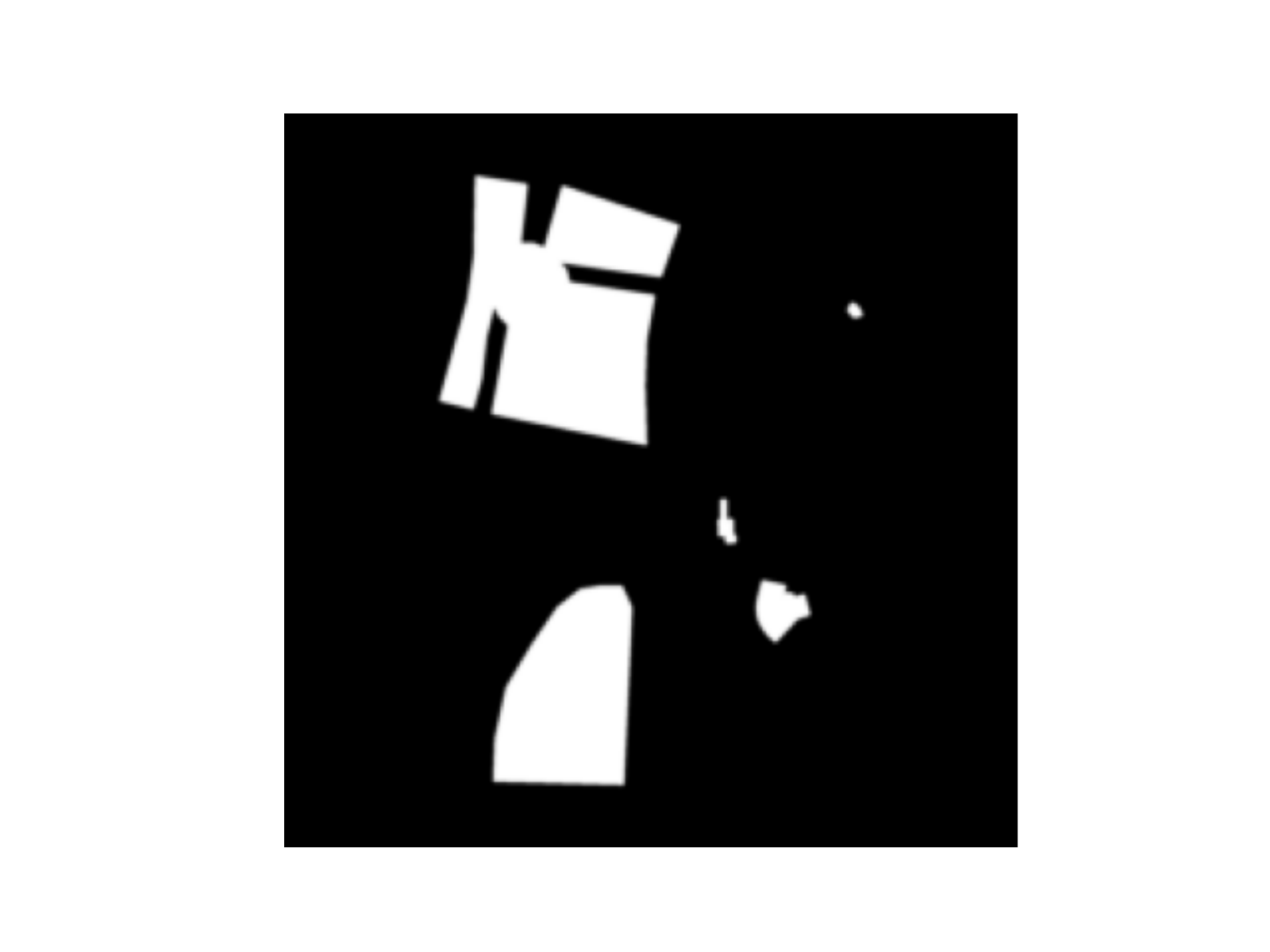}
         \caption{Channel 1}
         \label{fig:unet-input1}
     \end{subfigure}
     \hfill
     \begin{subfigure}[b]{0.19\textwidth}
         \centering
         \includegraphics[width=\textwidth,trim={3cm 1cm 3cm 1cm},clip]{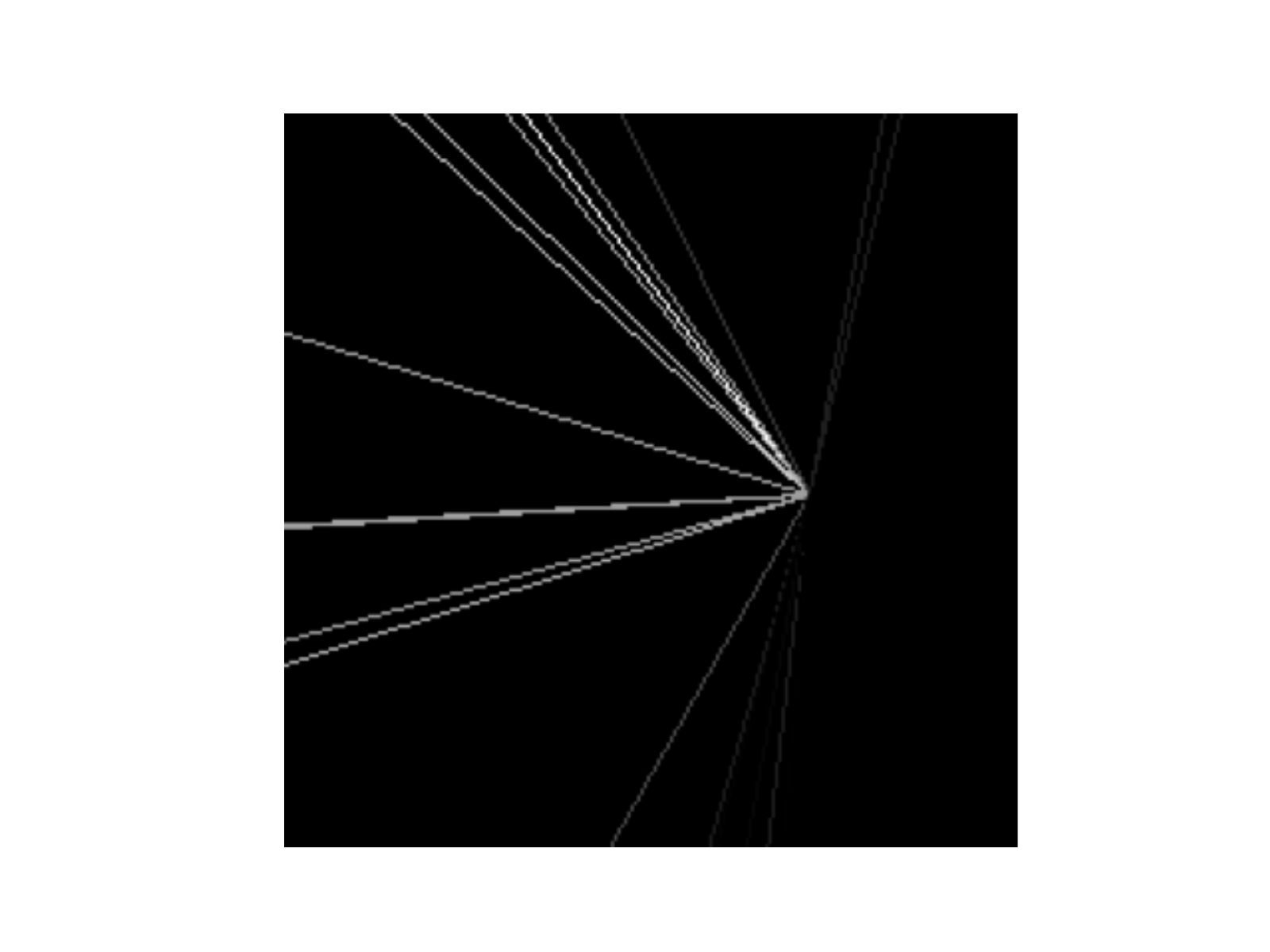}
         \caption{Channel 2}
         \label{fig:unet-input2}
     \end{subfigure}
     \hfill
     \begin{subfigure}[b]{0.19\textwidth}
         \centering
         \includegraphics[width=\textwidth,trim={3cm 1cm 3cm 1cm},clip]{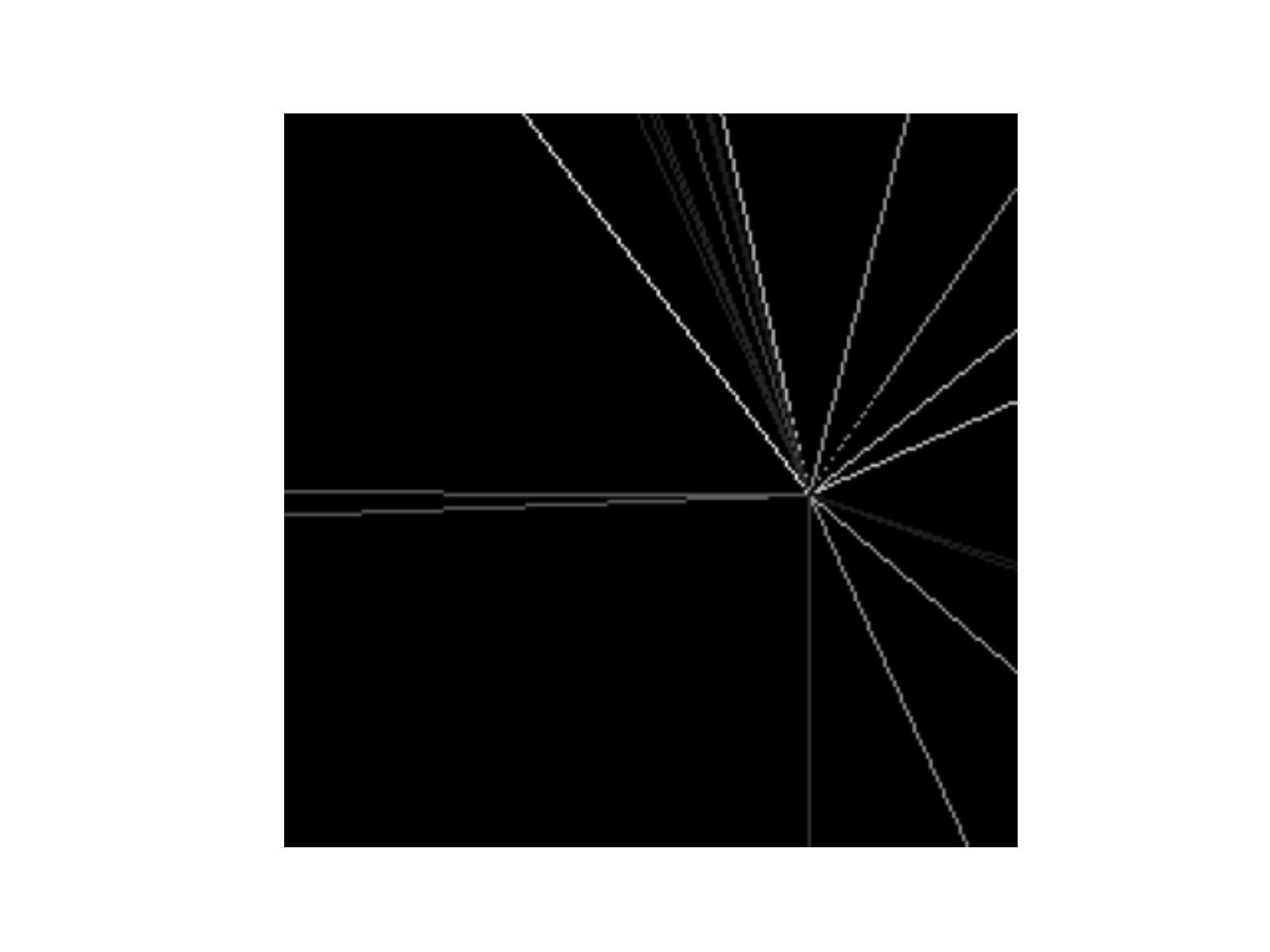}
         \caption{Channel 3}
         \label{fig:unet-input3}
     \end{subfigure}
     \hfill
     \begin{subfigure}[b]{0.19\textwidth}
         \centering
         \includegraphics[width=\textwidth,trim={3cm 1cm 3cm 1cm},clip]{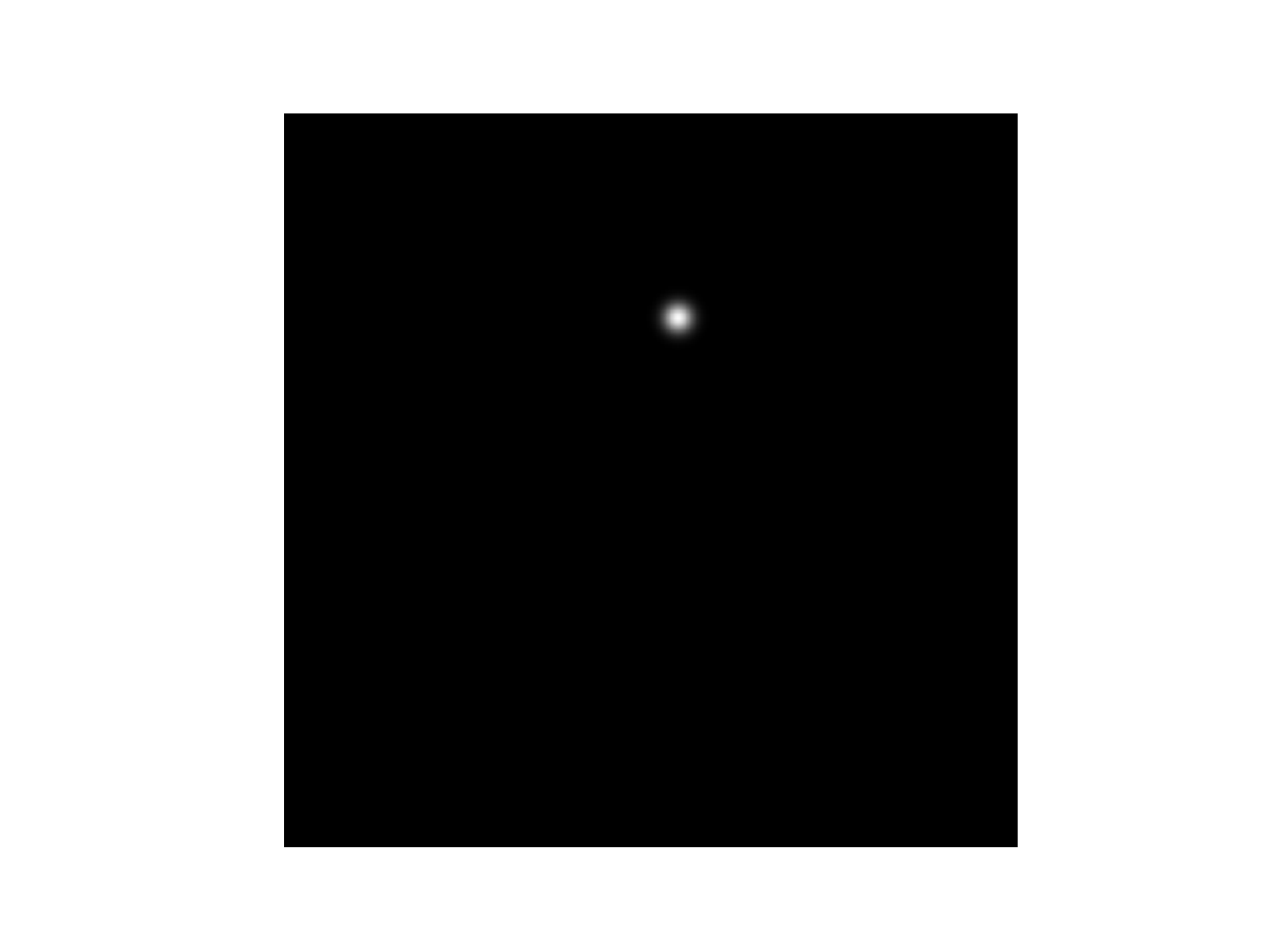}
         \caption{Output}
         \label{fig:unet-output}
     \end{subfigure}
        \caption{Input and output representations for UNet}
        \label{fig:unet}
\end{figure*}

In order to tackle the localization problem, we employ ML-driven approaches. 

\subsection{A baseline method with vector inputs}

A baseline ML method to approach the problem is a regular multi-layer perceptron (MLP) described in the codebase of the WAIR-D dataset \cite{WAIR-D}. It takes as an input the following 5-dimensional vector: $\left(w_{j,x}, w_{j,y}, \tau_{i,j,1}, \psi_{i,j,1}, \phi_{i,j,1} \right)$. Note that this approach takes only the shortest path of the radio link, where the length of the path is determined by $\tau_{i,j}$. Also it does not use map information. The network consists of three layers of sizes 256, 1024 and 256, and an additional output layer which produces the $x$ and $y$ coordinates of the source node. Root mean square is chosen as the loss function.

\subsection{A CNN with matrix inputs}

To leverage the other radio paths which are available but not used by the first approach, we have designed a simple convolutional neural network (CNN), the purpose of which is to broaden the utilization range of the available inputs. Its input is a matrix of size $(K_{i,j}, 5)$, where each row contains the parameters of each radio path. The rows are sorted by increasing order of $\tau_{i,j, k}$. The network consists of 3 layers of 1-dimensional convolutions with kernel size 1$\times$5 (these are designed to process each individual path), and another 3 layers of 2-dimensional convolutions with kernel size 3$\times$3. Then the outputs are pooled across the axis of radio paths to obtain a fixed sized representation of the entire radio link. A final linear layer maps the representation into two neurons to regress the coordinates of the user equipment (\cref{fig:arch-cnn}). Note that when $K_{i,j}=1$ this convolutional network collapses to an MLP with six intermediate layers. 

\subsection{A CNN with image inputs}

\begin{figure*}
    \centering
     
     \begin{subfigure}[b]{0.49\textwidth}
         \centering
         \includegraphics[width=\textwidth]{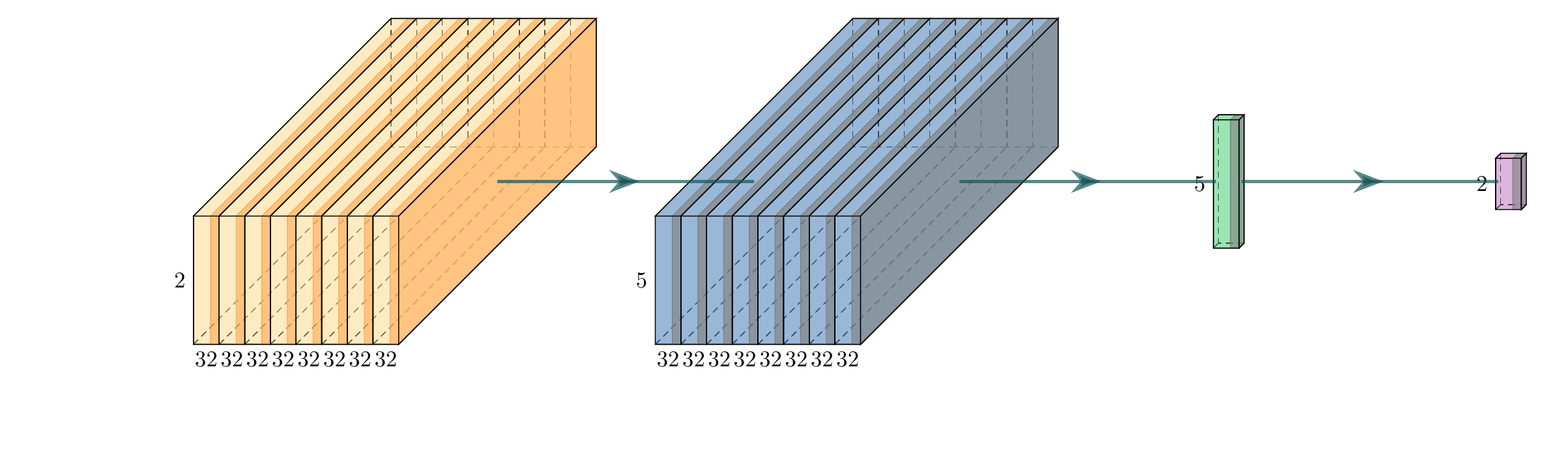}
         \caption{Matrix-based CNN}
         \label{fig:arch-cnn}
     \end{subfigure}
     \hfill
     \begin{subfigure}[b]{0.49\textwidth}
         \centering
         \includegraphics[width=\textwidth]{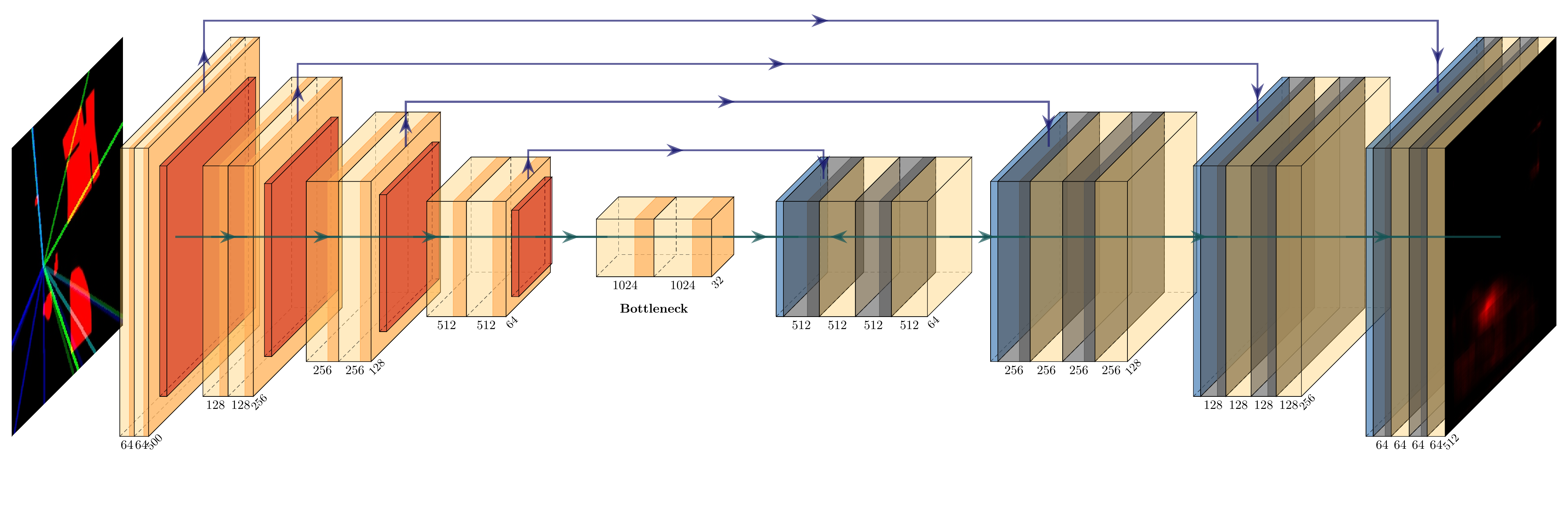}
         \caption{Image-based CNN (UNet)}
         \label{fig:arch-unet}
     \end{subfigure}
    \label{fig:architectures}
    \caption{CNN-based neural architectures defined in \cref{sec:models}.}
\end{figure*}

Finally, we designed another CNN for a fundamentally different input representation. We encode each sample of the input data (e.g. \cref{fig:input-sample}) as an image of size 224$\times$224 with three channels. The first channel represents the map $u_l$ encoded as a binary matrix, where the locations of the obstacles are encoded by value ``1'' (example in Fig. \ref{fig:unet-input1}). In the case that the maps have different sizes (such as in the  WAIR-D dataset \cite{WAIR-D}), we pad them to become a square and resize them to 224$\times$224 pixels. The second channel encodes the radio paths arriving at the base station, while the third channel encodes the radio paths departing from the source. Each radio path is depicted as a ray starting from the base station's location and going in the direction of angle of arrival (in the second channel, Fig. \ref{fig:unet-input2}) and angle of departure (in the third channel, Fig. \ref{fig:unet-input3}). We use the base location as the starting point of rays in the third channel because the location of the source is not known. We use UNet neural architecture \cite{unet} applied on this input (\cref{fig:arch-unet}). The output of the UNet is another single-channel map of size 224$\times$224px with pixel-wise sigmoid activation (to ensure that the output of each pixel is a number between 0 and 1). The output of each pixel is interpreted as a probability of the source node being at that location.


To train the network, we define the target of the neural network as a single-channel image of size 224$\times$224px where the location of the source user equipment is marked with a Gaussian kernel (Fig. \ref{fig:unet-output}). The size of the kernel is a hyperparameter. We use per-pixel binary cross-entropy loss. Additionally we experimented with dice loss \cite{diceloss} as an additional term in our loss function.

\section{Performance evaluation} \label{sec:perf}

In this section we present the performance evaluation results that we obtaining by using a synthetic dataset of wireless traces.

\subsection{The synthetic dataset}

The WAIR-D dataset~\cite{WAIR-D} contains 10.000 urban environments with scattered buildings based on maps of real cities, the locations of multiple base stations and source nodes, along with radio link parameters. 
We followed the ``Scenario 1'' subset of the data where each urban environment contains $|V_l| = 5$ base stations and $|W_l| = 30$ source nodes. This produces $150$ (base station, source node) pairs per environment. We used $L = 9.400$ maps as the training set, 500 as the validation set and 100 as the test set. 30\% of the 1.410.000 training pairs and 29\% of the 15.000 test pairs correspond to a NLOS scenario. 


\subsection{Benchmark}

We evaluate the prediction models by root mean square error (RMSE) of the predicted locations in meters. Additionally, we measure accuracy @$X$ meters which is defined as the ratio of samples for which the predicted location was within $X$ meters of the true location. All metrics are also reported on the subsets of LOS and NLOS samples. 

\subsection{Implementation details}
All the models were trained using Adam optimizer \cite{adam} with base learning rate $3\cdot 10^{-4}$. The MLP model and CNNs with the shortest and the two shortest paths were trained with batch size of 500. The batch size for the CNN using all the paths and UNet was 128. UNet was trained for 15 epochs on a single NVIDIA A100 GPU.

\subsection{Results}

\begin{figure}[t!]
    \centering
    \includegraphics[width=\columnwidth]{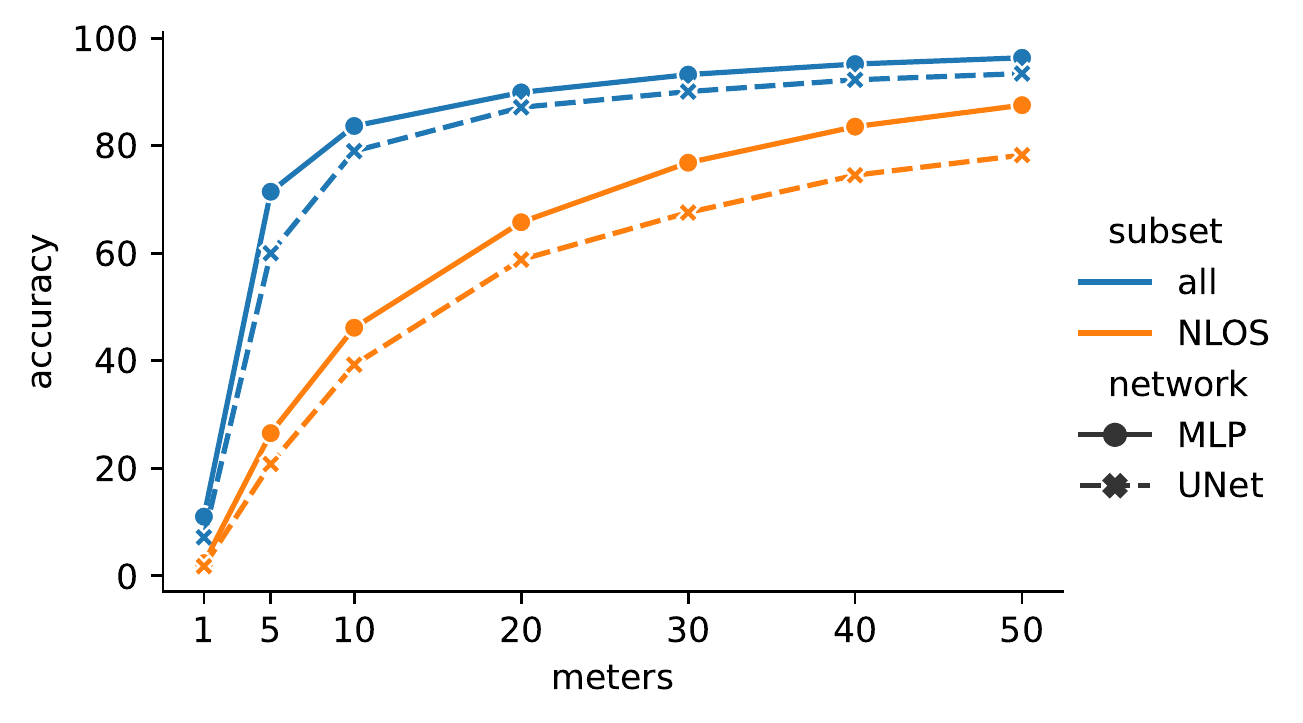}
    \caption{Accuracy of the MLP and UNet models with respect to the maximum tolerated error in meters.}
    \label{fig:accuracy-per-distance}
\end{figure}

We trained and compared the performance of the three models presented in section \ref{sec:models}. We focused our experimental analysis on obtaining insights about the LOS and NLOS cases separately. Table \ref{tab:main-results} shows the performance of the three models in terms of RMSE and accuracy within 10 meters. The performance in terms of accuracy of the predicted location within some distance threshold is shown in Fig. \ref{fig:accuracy-per-distance} for various thresholds. The MLP method works quite well for LOS samples (99\% 10 meter accuracy)  but deteriorates on NLOS cases (46\%). This is expected, as it uses the information only from the shortest path of the radio link.

Surprisingly, the matrix-based CNN designed to leverage all available radio paths of a given radio link performs much worse compared to the MLP which uses just the shortest radio path. To ensure that the reason of the poor performance is not in the architecture of the neural network, we trained two additional versions of the CNN: with the shortest path only, and with the two shortest paths only. The single-path version performs similarly to the MLP model. The inclusion of the second path already worsens the performance (Table~\ref{tab:main-results}). We hypothesize that the additional radio paths provide spurious signals that cause the neural network to quickly overfit on the training set.

\begin{table}[t!]
\begin{tabular}{@{}lrrrrrr@{}}
\toprule
                       & \multicolumn{3}{c}{RMSE}                                                     & \multicolumn{3}{c}{Acc @10M}                                                 \\
\textit{Subset}        & \multicolumn{1}{c}{All} & \multicolumn{1}{c}{LOS} & \multicolumn{1}{c}{NLOS} & \multicolumn{1}{c}{All} & \multicolumn{1}{c}{LOS} & \multicolumn{1}{c}{NLOS} \\ \midrule
MLP (shortest path)    & 8.4                     & 2.7                     & 22                     & 84\%                    & 99\%                    & 46\%                     \\
CNN (shortest path)    & 9.5                     & 4.1                     & 22.4                        & 81\%                    & 96\%                    & 44\%                     \\
CNN (2 shortest paths) & 9.6                     & 4.2                     & 22.5                     & 81\%                    & 96\%                    & 44\%                     \\
CNN (all paths)        & 38.5                      & 36.5                      & 41.8                       & 15\%                    & 17\%                    & 11\%                     \\
UNet (all paths)       & 13.8                    & 4.5                     & 36.6                     & 78\%                    & 93\%                    & 39\%                     \\ \bottomrule
\end{tabular}
\caption{The performance of our models on the test set of the WAIR-D dataset \cite{WAIR-D}.}
\label{tab:main-results}
\end{table}

\begin{figure}[t!]
    \centering
    \includegraphics[width=\columnwidth]{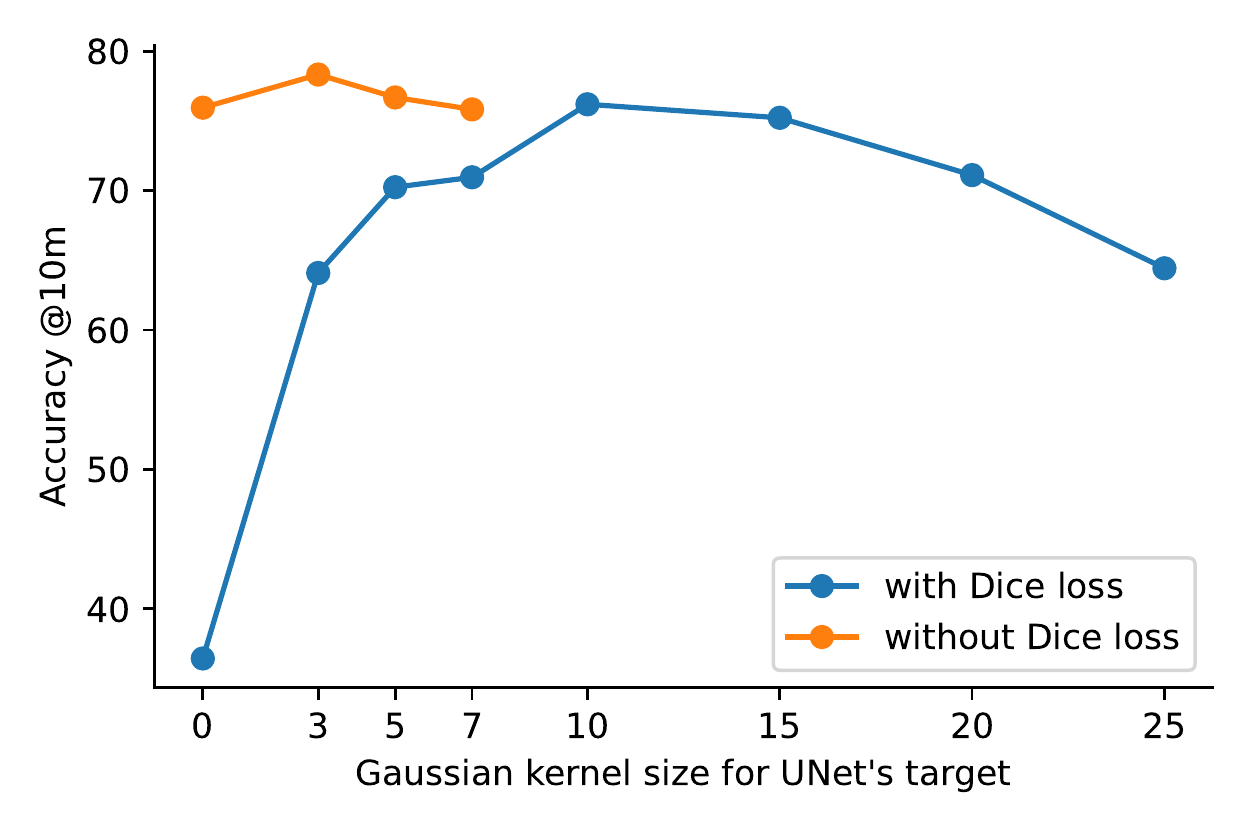}
    \caption{Accuracy of the UNet model with various hyperparameters on the validation set of WAIR-D dataset.}
    \label{fig:unet-hyperparams}
\end{figure}

We trained multiple UNets with varying kernel sizes with and without dice loss. As seen in Fig. \ref{fig:unet-hyperparams}, dice loss works better with larger kernel sizes. Still, the best performance on the validation set was achieved using kernel size 3 and without dice loss. This particular model was used for all further analysis.

\begin{figure*}[t!]
     \centering
     \begin{subfigure}[b]{0.49\textwidth}
         \centering
         \includegraphics[width=\textwidth,trim={0 3cm 0 3cm},clip]{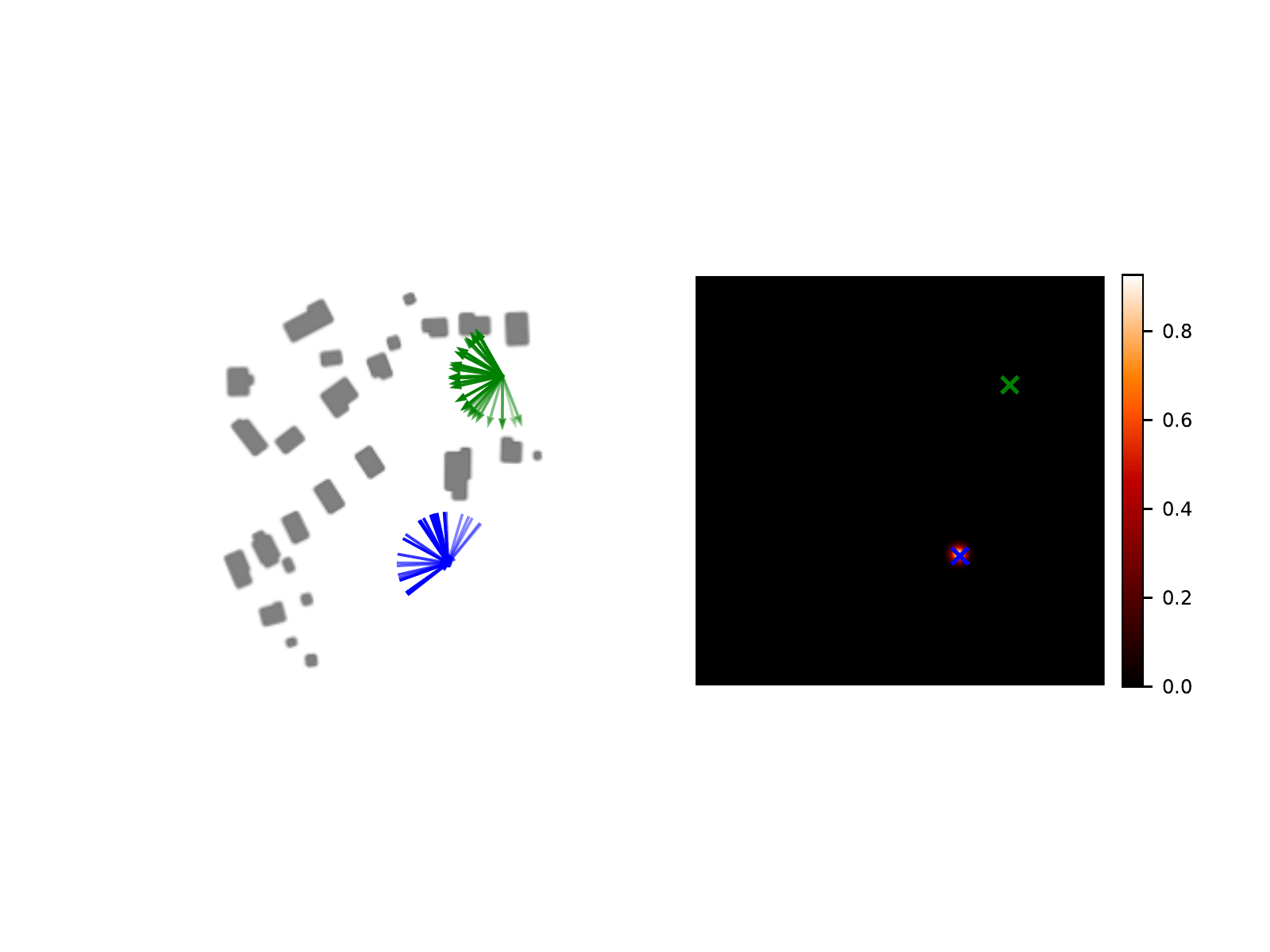}
         \caption{\#11000}
         \label{fig:sample-11000}
     \end{subfigure}
     \hfill
     \begin{subfigure}[b]{0.49\textwidth}
         \centering
         \includegraphics[width=\textwidth,trim={0 3cm 0 3cm},clip]{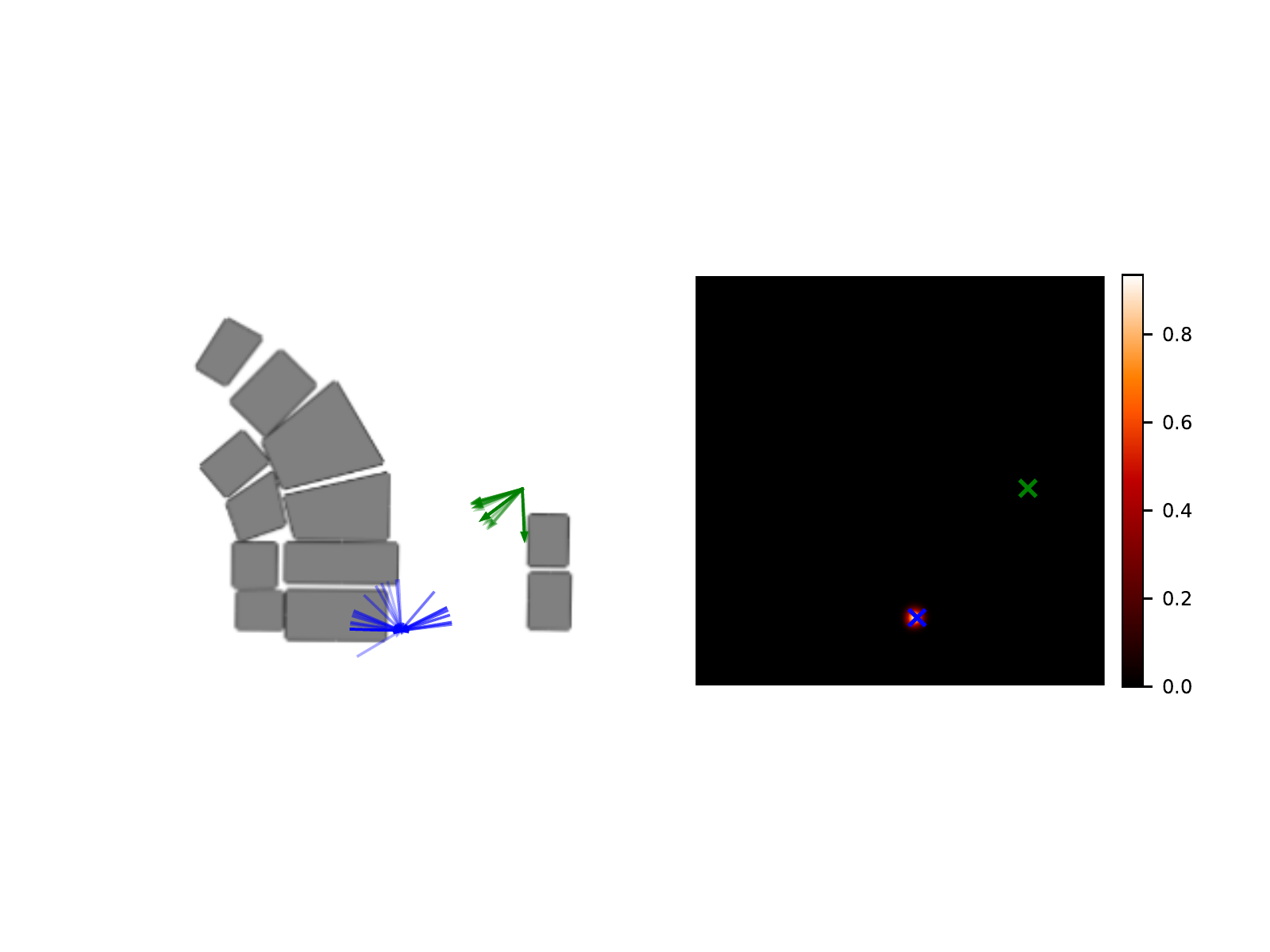}
         \caption{\#7000}
         \label{fig:sample-7000}
     \end{subfigure}
     \hfill
     \begin{subfigure}[b]{0.49\textwidth}
         \centering
         \includegraphics[width=\textwidth,trim={0 3cm 0 3cm},clip]{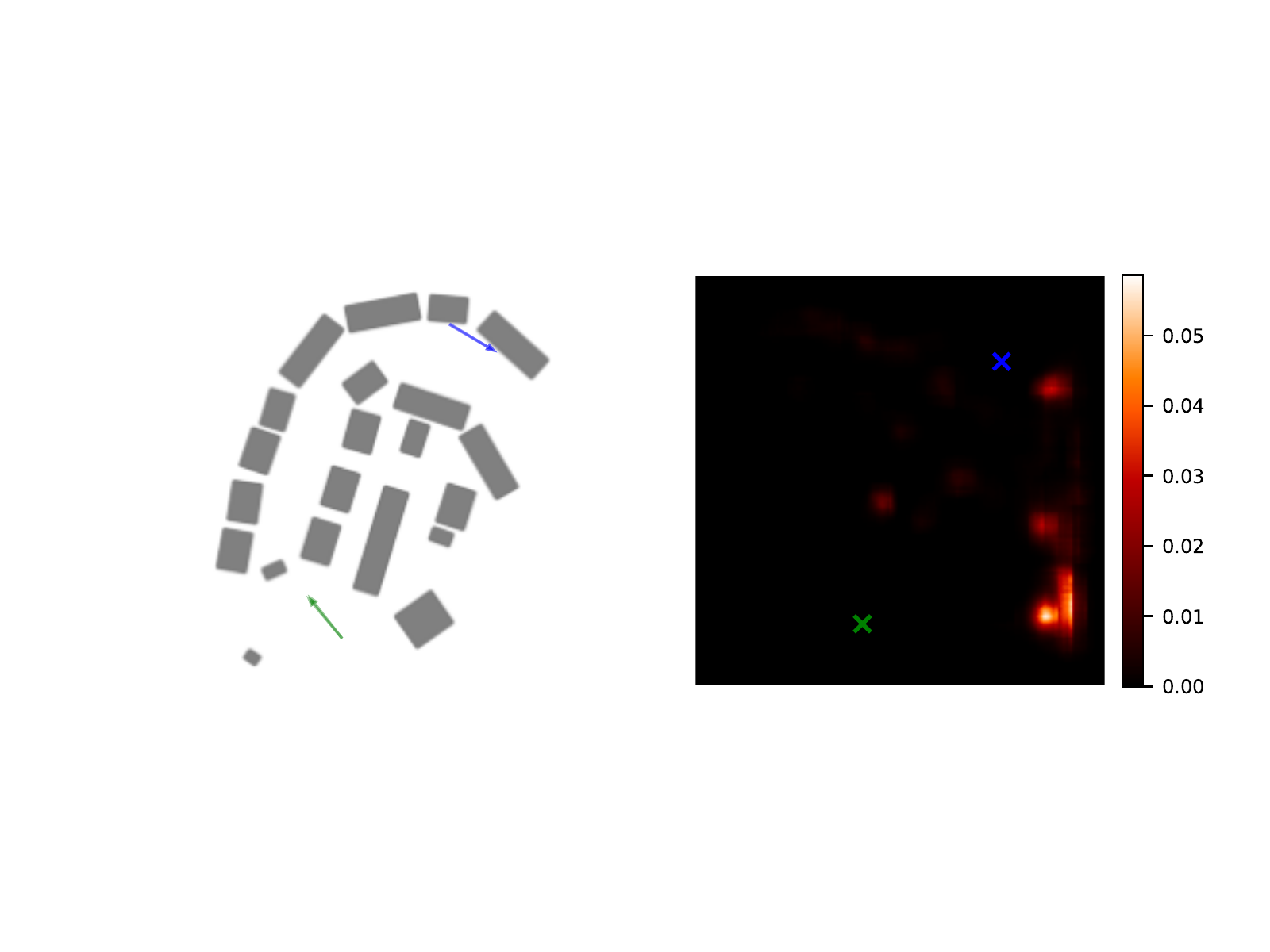}
         \caption{\#4000}
         \label{fig:sample-4000}
     \end{subfigure}
     \hfill
     \begin{subfigure}[b]{0.49\textwidth}
         \centering
         \includegraphics[width=\textwidth,trim={0 3cm 0 3cm},clip]{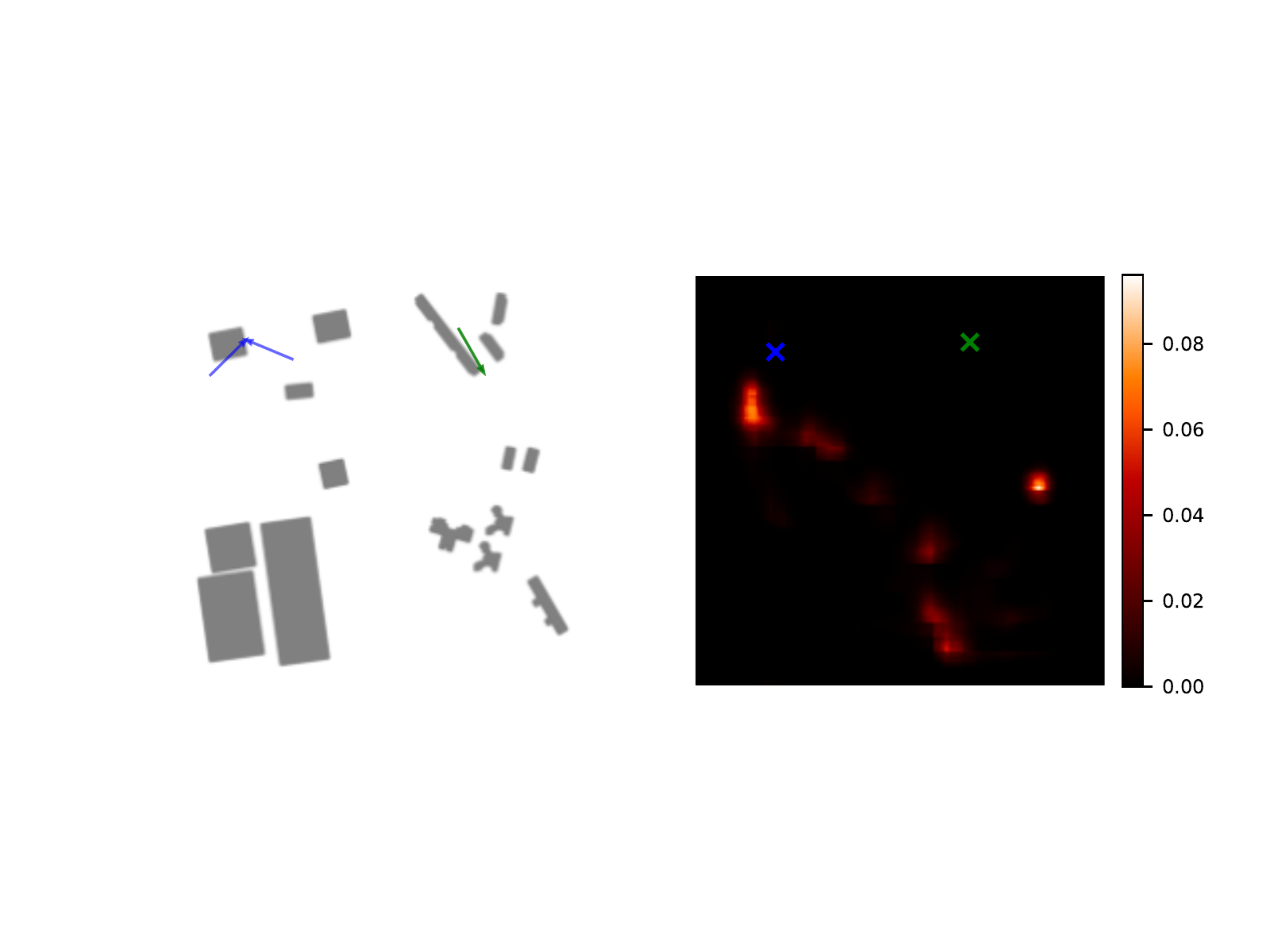}
         \caption{\#5000}
         \label{fig:sample-5000}
     \end{subfigure}
     \hfill
     
     \caption{Random input samples along with UNet predictions.}
     \label{fig:unet-predictions}
\end{figure*}

\begin{table}[]
\centering
\begin{tabular}{@{}crrr@{}}
\toprule
Max tolerated error & \multicolumn{1}{c}{Top 1 acc.} & \multicolumn{1}{c}{\parbox[t]{1.5cm}{Top 5 acc.\\ ($\text{GMM}_{\text{mean}}$)}} & \multicolumn{1}{c}{\parbox[t]{1.5cm}{Top 5 acc.\\ ($\text{GMM}_{\text{argmax}}$)}} \\ \midrule
10 meters               & 39.4\%         & 41.1\%         & 40.7\% \\
20 meters               & 60\%           & 62.4\%         & 62.2\% \\
30 meters               & 67.7\%         & 73.1\%         & 72.1\% \\
40 meters               & 74.7\%         & 80.3\%         & 80.2\% \\
50 meters               & 78.5\%         & 85\%           & 84.3\% \\ \bottomrule
\end{tabular}
\caption{The difference between Top 1 and Top 5 accuracy measures for NLOS samples with respect to different thresholds of tolerated errors.}
\label{tab:gmm}
\end{table}

The best UNet method leverages all radio paths and the map of the environment, and gets close to the performance of the MLP (39\% vs. 42\% accuracy at 10 meters for NLOS cases). While it failed to surpass the MLP's performance on NLOS samples, it gives a rich output which enables more nuanced interpretation of the prediction.

Fig. \ref{fig:unet-predictions} shows randomly selected input samples along with the predictions given by UNet. Whenever the user equipment is within LOS of the base station, the prediction of the model is well concentrated around one point which is usually quite close to the ground truth location (\cref{fig:sample-11000,fig:sample-7000}). \Cref{fig:sample-4000,fig:sample-5000} depict NLOS cases when the model has trouble in predicting the correct location. Instead it gives non-zero probability to many regions of the map. On one hand this can be interpreted as predicting multiple candidates of possible locations of the device, which might be leveraged in some applications (\cref{sec:gmm}). Another interpretation is that the model is uncertain about its predictions and we might want to discard the lowest confidence predictions (\cref{sec:uncertainty}).

\begin{figure*}[t!]
    \centering
    \begin{subfigure}[b]{0.48\textwidth}
         \centering
         \includegraphics[width=\textwidth,trim={0 3cm 0 3cm},clip]{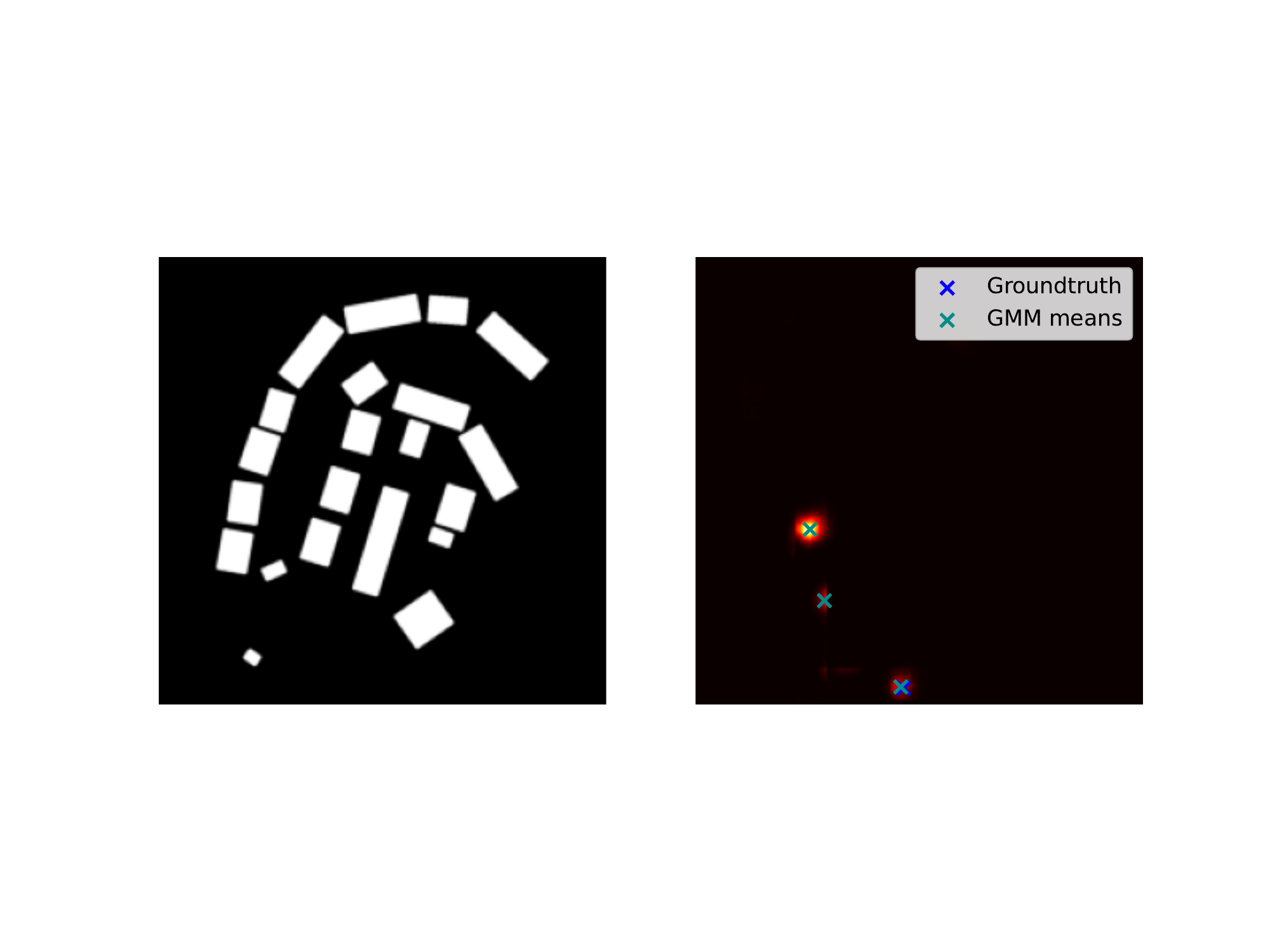}
         \caption{An example of top three locations detected by $\text{GMM}_{\text{mean}}$ on a prediction heatmap. One of the locations is quite accurate.}
         \label{fig:gmm-sample}
     \end{subfigure}
     \hfill
    \begin{subfigure}[b]{0.48\textwidth}
         \centering
         \includegraphics[width=\textwidth,trim={0 3cm 0 3cm},clip]{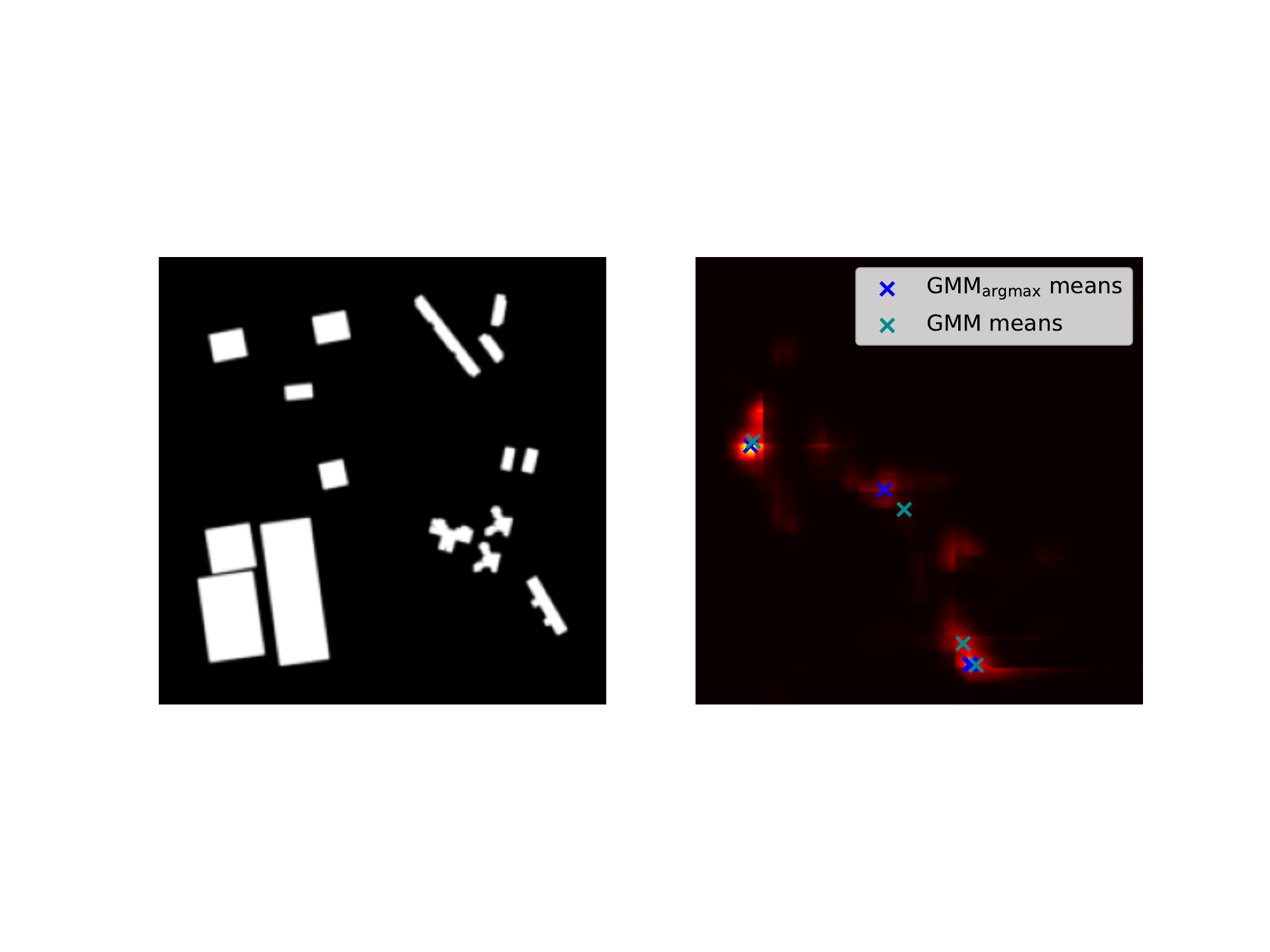}
         \caption{An example where $\text{GMM}_{\text{mean}}$ predicts a location that does not coincide with the location having the highest score.}
         \label{fig:gmm-mean-problem}
     \end{subfigure}
     \hfill
     
    \caption{Fitting Gaussian Mixture Model on the prediction heatmaps.}
    \label{fig:gaussian-plot}
\end{figure*}

\begin{figure*}[t!]
    \centering
     \begin{subfigure}[b]{0.48\textwidth}
         \centering
         \includegraphics[width=\textwidth]{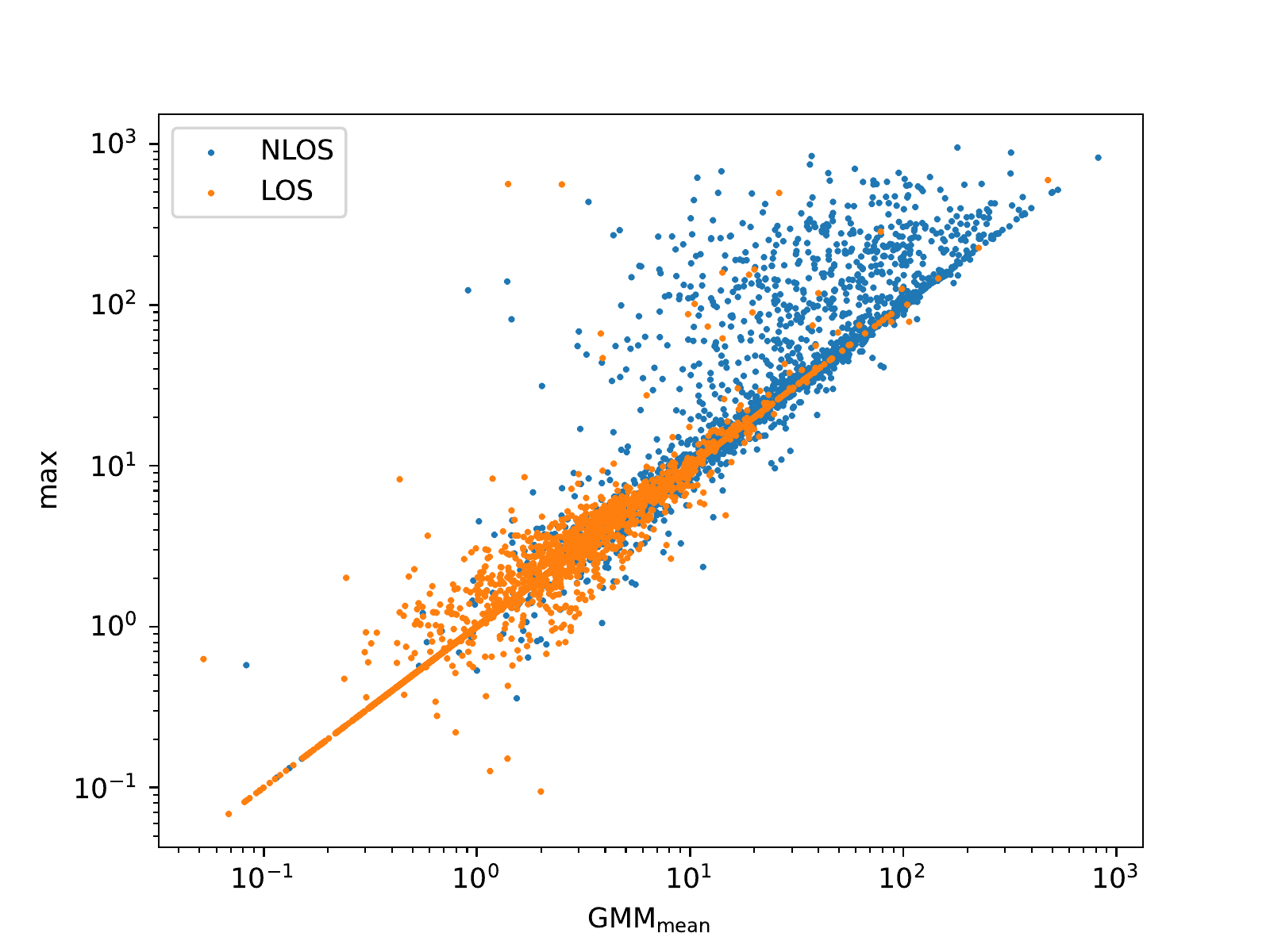}
         \caption{x-axis is the error of the location predicted by $\text{GMM}_{\text{mean}}$}
         \label{fig:gmm-scatter}
     \end{subfigure}
     \hfill
     \begin{subfigure}[b]{0.48\textwidth}
         \centering
         \includegraphics[width=\textwidth]{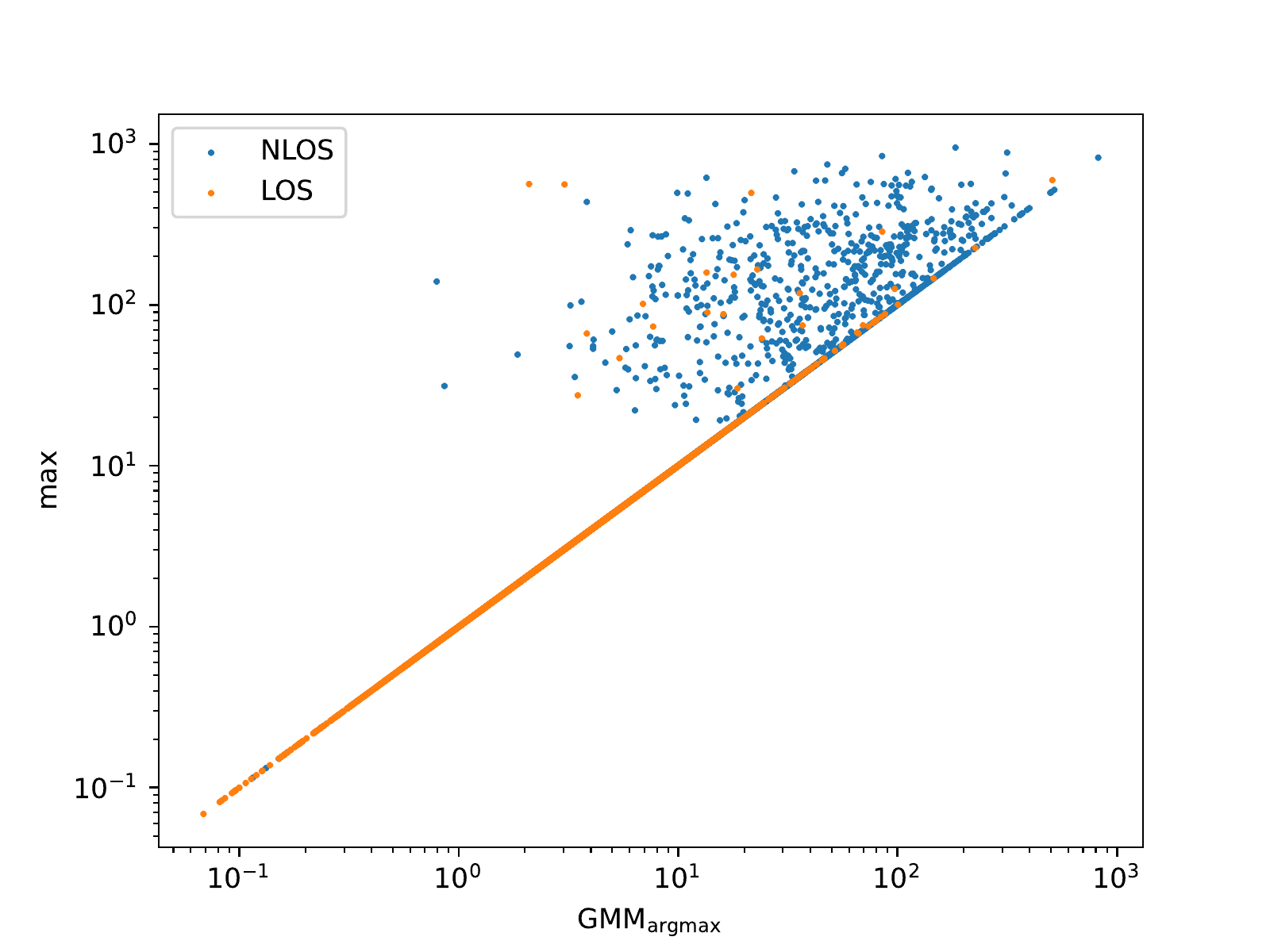}
         \caption{x-axis is the error of the location predicted by $\text{GMM}_{\text{argmax}}$.}
         \label{fig:gmm-scatter-argmax}
     \end{subfigure}
     \hfill
    \caption{Scatter plots of errors of UNet predictions. Each point corresponds to one input sample. y-axis is the error in meters of the most probable location, x-axis is the error of the closest prediction among the top 5 predictions given by the GMMs. }
    \label{fig:gaussian-scatter}
\end{figure*}

\subsection{Predicting top five locations}\label{sec:gmm}
In some applications when it is hard to locate the source immediately, one might want to predict a set of $M$ candidate locations. For example, if the transmitter requires medical help, and the rescue team cannot find the transmitter in the top predicted location, they might want to learn about other candidate locations where they can look into. This is impossible to achieve with machine learning models that are trained to regress the location coordinates directly. 

To get the top $M$ predicted locations of the UNet model we fit a Gaussian mixture model on the prediction heatmap for each input sample with $M$ kernels (Fig.~\ref{fig:gmm-sample}). To determine $M$, we fit $1,2,\ldots$ kernels and measure the distance between the closest modes. If the distance between the closest pair for $m$ kernels is less than some threshold, we choose $M = m-1$. We do not take more than $5$ kernels. Thus we obtain up to five means of Gaussians as the top predictions. We measure the distance between the ground truth location and the closest prediction. Fig.~\ref{fig:gmm-scatter} shows that for many NLOS samples the error of the closest prediction among top $5$ candidates (horizontal axis) is better than the error of the top 1 prediction (vertical axis). For LOS samples there is almost no difference, which implies that the prediction heatmaps presented in the first row of Fig.~\ref{fig:unet-predictions} are quite representative for LOS scenario. We call this approach under $\text{GMM}_{\text{mean}}$.

We also notice that some of the points on \cref{fig:gmm-scatter} are below the diagonal, which means the closest among the top five predictions is worse than the top prediction. This happens because sometimes the mean of the kernel does not coincide with the highest scoring location in its neighborhood. As another alternative, for each kernel we picked the highest scoring location within some distance from the kernel mean. We call this approach $\text{GMM}_{\text{argmax}}$ and show its performance on \cref{fig:gmm-scatter-argmax}. Table \ref{tab:gmm} shows top-five accuracy for both approaches. Accuracy of NLOS samples slightly increases across all thresholds. We see that the increase is more significant for larger thresholds. While $\text{GMM}_{\text{argmax}}$ performs worse on average than $\text{GMM}_{\text{mean}}$, there are no samples for which $\text{GMM}_{\text{argmax}}$ is worse than the top-one prediction.

\subsection{Identifying Incorrect Predictions}\label{sec:uncertainty}
As the prediction of the UNet network is a heatmap over the entire map, we can use the heatmap characteristics to measure the uncertainty of predictions. We propose measuring the uncertainty by the dispersion of predictions on the heatmap. Formally, we count the number of pixels with an intensity greater than half the intensity of the highest prediction.

\begin{table}[t!]
\centering
\begin{tabular}{@{}lrrrrrrrr@{}}
\toprule
                  & \multicolumn{4}{c}{MLP}                                                                                 & \multicolumn{4}{c}{UNet (kernel size=3)}                                                                \\ \midrule
                  & \multicolumn{2}{c}{RMSE}                           & \multicolumn{2}{c}{Acc. @10m}                  & \multicolumn{2}{c}{RMSE}                           & \multicolumn{2}{c}{Acc. @10m}                  \\
                  & \multicolumn{1}{c}{All} & \multicolumn{1}{c}{NLOS} & \multicolumn{1}{c}{All} & \multicolumn{1}{c}{NLOS} & \multicolumn{1}{c}{All} & \multicolumn{1}{c}{NLOS} & \multicolumn{1}{c}{All} & \multicolumn{1}{c}{NLOS} \\\midrule
Base model        & 9                       & 24                       & 82\%                    & 42\%                     & 14                      & 36                       & 77.5\%                  & 38.6\%                   \\
- w/o map &                         &                          &                         &                          & 20                      & 60                       & 71.5\%                  & 18.7\%                   \\
- w/o AoD     & 16                      & 43                       & 68\%                    & 18\%                     & 24                      & 67                       & 69.9\%                  & 22.6\%                   \\
- w/o AoA     & 21                      & 54                       & 61\%                    & 17\%                     & 31                      & 89                       & 65.7\%                  & 16.0\%                   \\ \bottomrule
\end{tabular}
\caption{Ablation analysis on MLP and UNet models with respect to input variables. Numbers reported on the test set.}
\label{tab:ablation}
\end{table}

\begin{figure}[t!]
    \centering
    \begin{subfigure}[b]{0.85\columnwidth}
         \centering
         \includegraphics[width=0.9\textwidth]{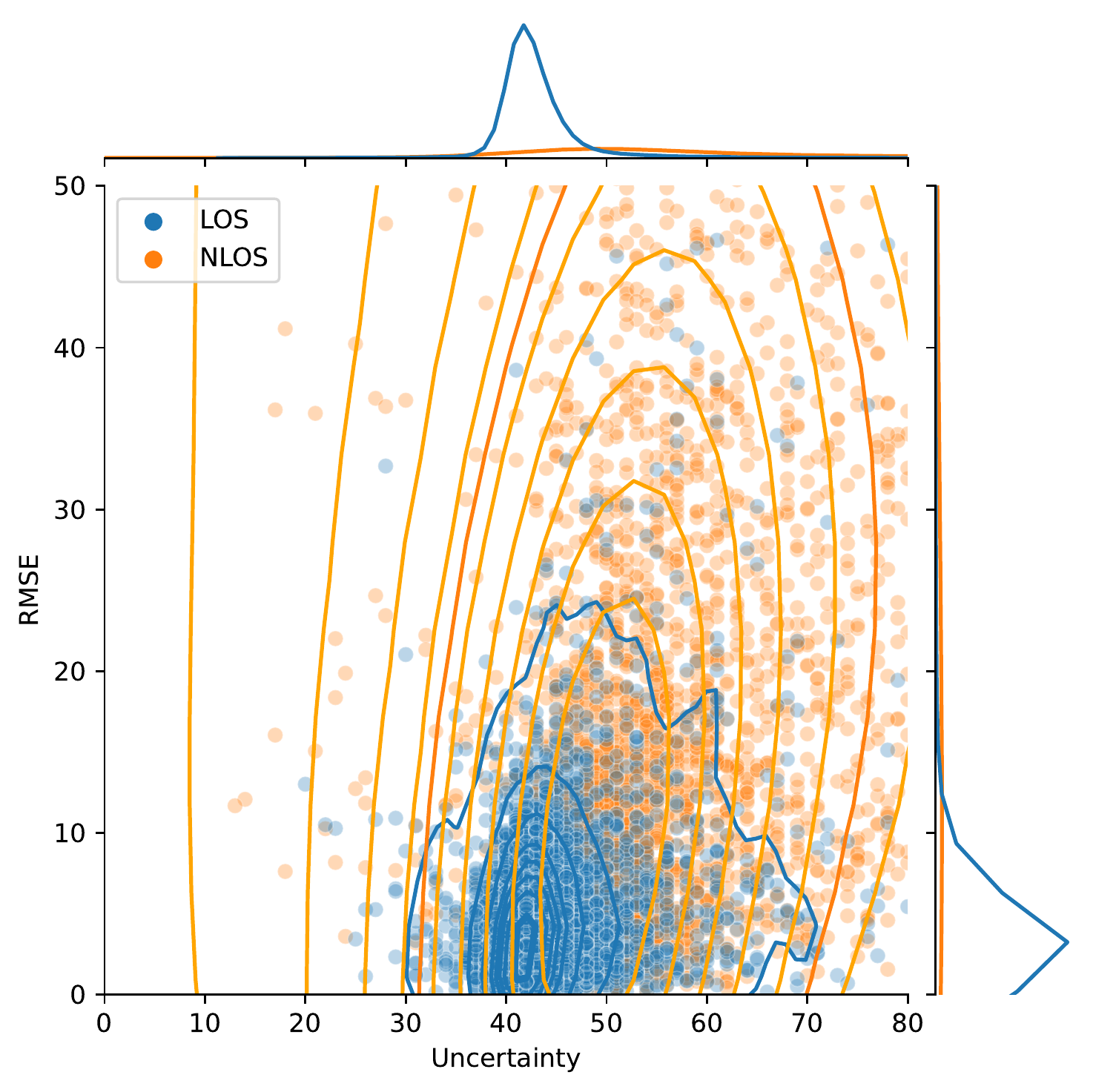}
         \caption{The relationship between prediction confidence (as measured by the number of relatively ``active'' pixels in the heatmap) and RMSE. Each dot is one input sample.}
         \label{fig:confidence-scatter}
     \end{subfigure}
     
     \begin{subfigure}[b]{0.85\columnwidth}
         \centering
         \includegraphics[width=\textwidth]{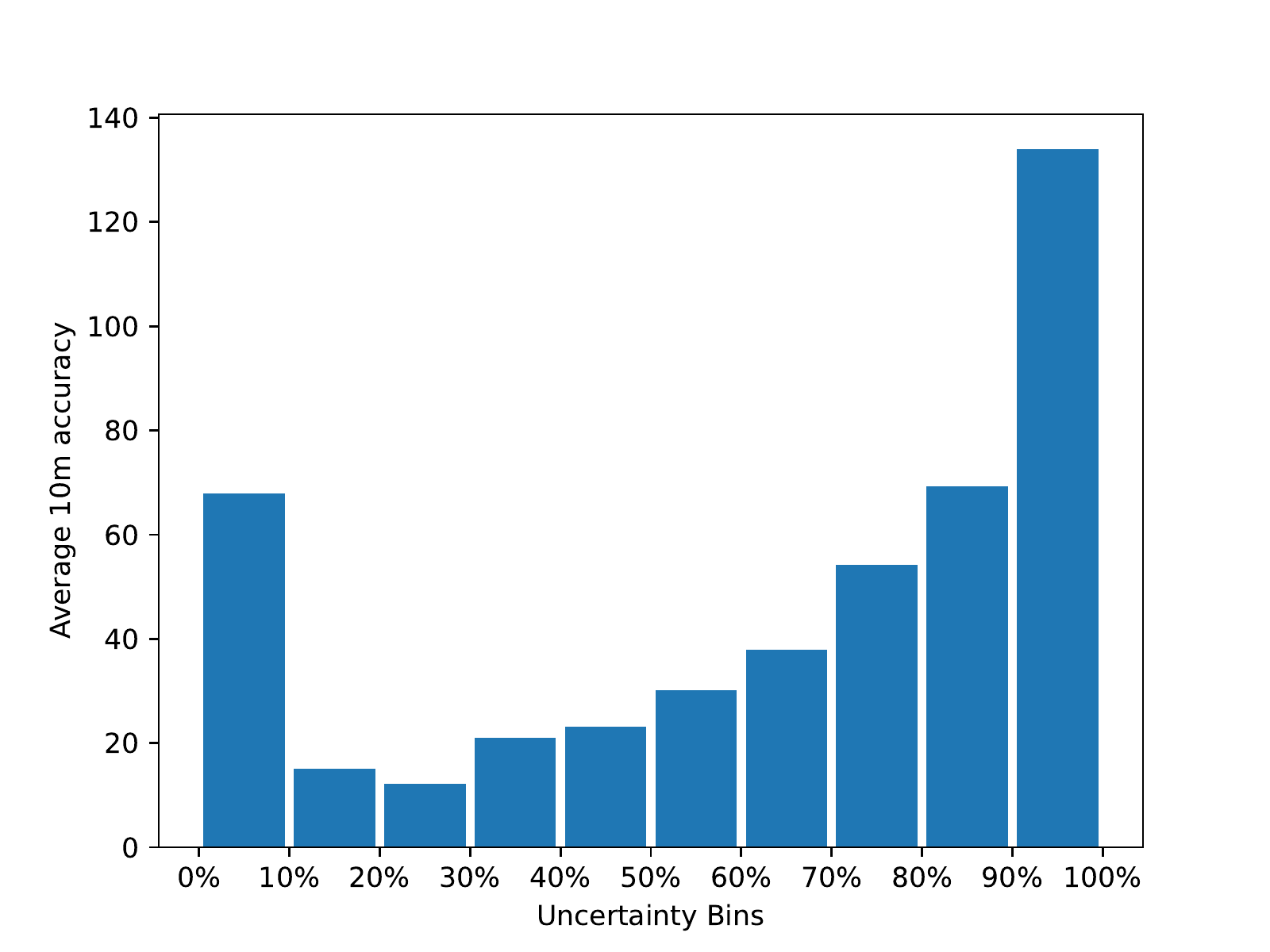}
         \caption{RMSE of the UNet model across the ten uncertainty bins of equal size (NLOS samples only).}
         \label{fig:confidence-bins-rmse}
     \end{subfigure}
     
    \caption{Uncertainty of the predictions correlates well with the errors on NLOS samples.}
    \label{fig:confidence}
\end{figure}

Fig.~\ref{fig:confidence-scatter} shows the relationship between our measure of uncertainty (horizontal axis) and RMSE (vertical axis). LOS samples have quite low uncertainty and low errors. NLOS samples have huge errors and high uncertainty with some noticeable correlation. In Fig.~\ref{fig:confidence-bins-rmse}, we group the predictions into equal-sized 10\% bins based on our uncertainty measure and compute the prediction error (RMSE) for each bin. It is evident that the bins with higher uncertainty have substantially greater errors compared to those with lower uncertainty. In practice, this will allow the user to ignore the uncertain predictions made by the model and achieve lower prediction error for the more certain subset of the predictions.

\subsection{Ablation over input variables}

For both MLP and UNet we measured the impact of input variables by performing experiments with some variables removed. As shown in Table \ref{tab:ablation}, the biggest drop in performance from the base UNet model in terms of overall accuracy (with a 10m tolerance threshold) is observed when removing the angles of arrival. For the NLOS samples the map is almost as important as AoA.

For MLP models, removing AoD and AoA have similar impact on the accuracy for NLOS samples, but for the entire dataset (including LOS samples) AoA is more important. This is due to the fact that, in realistic application cases where the source node is completely unknown or even malicious, the base station cannot rely on information received from the source node, as it might be erroneous \cite{4384492}. Therefore, application designers (on a per case basis), can consider relying more on AoA than on AoD, as the former can be more accessible and reliable.  In terms of average RMSE, the behavior is similar.

\section{Conclusion} \label{sec:conc}


In this paper we explored the challenging scenario of single-source single-base wireless localization in non-line-of-sight scenario. We tested several machine learning methods on different representations of inputs, including a UNet-based deep neural network designed to produce rich outputs. We showed how this structure of outputs can be used to produce multiple candidates of source locations and to detect incorrect predictions given by the model. We showed that AoA information is more important than AoD in all machine learning methods we tried. These insights motivate future work on building better neural architectures and input representations to get improved performance of NLOS localization along with additional features that are useful in various practical applications. 

\section{Acknowledgement}
The work of H. Khachatrian, R. Darbinyan and R. Mkrtchyan was partly supported by the RA Science Committee grant No. 22rl-052 (DISTAL). The work of T. P. Raptis was partly supported by the European Union under the Italian National Recovery and Resilience Plan (NRRP) of NextGenerationEU, partnership on “Telecommunications of the Future” (PE00000001 - program “RESTART”). The authors would like to thank Karen Hambardzumyan and Vahan Huroyan for the insightful discussions. The authors would also like to thank BlueQubit for providing access to computational resources.

\balance

\bibliographystyle{plain}
\bibliography{references}

\end{document}